\documentclass[sigconf,10pt]{acmart}

\usepackage[utf8]{inputenc}
\usepackage{subfigure}
\usepackage{graphicx}

\PassOptionsToPackage{hyphens}{url}
\usepackage{hyperref}

\usepackage{pythonhighlight}

 \setcopyright{none} \renewcommand\footnotetextcopyrightpermission[1]{}

\begin{document}

\title{Are Home Security Systems Reliable?}

\author{Christopher~Vattheuer}
\affiliation{
\institution{University of Manitoba}
\country{Canada}}
\email{christophervatt@gmail.com}

\author{Charlie Liu}
\affiliation{
\institution{University of Waterloo}
\country{Canada}}
\email{charlie.liu@uwaterloo.ca}

\author{Ali Abedi}
\affiliation{
\institution{Stanford University}
\country{USA}}
\email{abedi@stanford.edu}

\author{Omid Abari}
\affiliation{
\institution{UCLA}
\country{USA}}
\email{omid@cs.ucla.edu}

\begin{abstract}
    Home security systems have become increasingly popular since they provide an additional layer of protection and peace of mind. 
    These systems typically include battery-powered motion sensors, contact sensors, and smart locks.
    Z-Wave is a very popular wireless communication technology for these low-power systems.
    In this paper, we demonstrate two new attacks targeting Z-Wave devices. First, we show how an attacker can remotely attack Z-Wave security devices to increase their power consumption by three orders of magnitude, reducing their battery life from a few years to just a few hours. Second, we show multiple Denial of Service (DoS) attacks which enables an attacker to interrupt the operation of security systems in just a few seconds. Our experiments show that these attacks are effective even when the attacker device is in a car 100 meters away from the targeted house.
\end{abstract}

\maketitle
\pagestyle{plain} 

\section{Introduction}
Home security systems have become very popular over the past few years. Today, more than 40\% of Americans own a home security system. Besides homes, many stores also use the same security systems. Multiple major technology companies (such as Amazon) as well as many start-ups are significantly investing in this domain and have introduced many different home security products. It is expected that the market for home security systems will reach 78.9 billion dollars by 2025\cite{Market2025}. 

A typical home security system consists of  motion detection and contact sensors installed on doors and windows.
To make their installation simple and easy, these devices are battery powered (with typical expected life-times of approximately 1-3 years), and utilize low-power wireless technologies for communicating to a base station. 
Z-Wave is the most popular wireless technology used in home security systems. In fact, 
9 out of 10 leading security companies use Z-Wave for their products, and there are over 3,300 Z-Wave certified products that are currently available in the market~\cite{ZwaveStats,ZwaveAlliance}.
Therefore, the security of the Z-Wave protocol is very crucial in the reliability of home security systems~\cite{homeSecurityReview}.

In this paper, we ask whether today's Z-Wave based home security systems are reliable. One na\"ive approach to attack these systems is to jam their wireless band, preventing them from communicating. However, this attack will not be very effective for multiple reasons. First, the attacker must be very close to the house and needs to transmit very high power RF signals toward the victim devices
Second, many home security systems 
detect interference and try to avoid it and notify their user about a potential attack~\cite{Jamming}.
We build on our recent study that has shown that all existing WiFi devices respond to fake packets transmitted to them~\cite{polite-wifi, polite-wifi-arxiv, wi-peep}.
In this paper, we show how an attacker can exploit vulnerabilities in the Z-Wave protocol to disable Z-Wave devices from up to 100 meters away, using a small off-the-shelf software radio.


We present two attacks against Z-Wave devices namely, battery draining and denial of service attacks.
We show that an attacker can completely drain the battery of home security systems by sending fake packets to a target device and forcing it to continuously transmit some response packets.
We show that this attack can increase the power consumption of Z-Wave devices by up to 1500 times!
We also present and investigate some denial of service attacks (DoS) against individual Z-Wave devices or their entire network.
We demonstrate that by sending some special packets to Z-Wave devices we can confuse them so that their operation is interrupted. During this attack security sensors stop functioning. For example, if someone passes in front of a motion sensor or if a door equipped with a contact sensor is opened, no alarm is activated nor any notification is sent to the user.
Finally, we propose some potential solutions to defend against these attacks. 
Although this paper mostly focuses on home security systems, the same attack can be performed on other devices, such as personal wearable,medical devices, etc., which use Z-Wave or similar technologies for communication. 


 In this paper, we make the following contributions:
 \begin{itemize}
    \item We design novel techniques for forcing Z-Wave devices out of the power saving mode to either interrupt their service or drain their batteries. Using these techniques, we show, for the first time, how an attacker can remotely reduce the battery life of Z-Wave devices from years to hours.
    
    \item We also design new Denial of Service attacks that deactivate Z-Wave security sensors, even when the attacker is 100 meters away from the victim device.
    \item We propose potential solutions to defend against these attacks, to make Z-Wave devices more secure.
    
 \end{itemize}

\section{Background}
\label{sec:background}
Nowadays, many smart home IoT devices use Z-Wave for communicating to a base station or hub. Z-Wave is  popular in devices that operate on small batteries and need to be very energy-efficient (such as door sensors). In this section, we provide some background on the aspects of the Z-Wave technology that are related to the attacks presented in this paper.

Z-Wave has recently become a common communication technology for smart home and security systems. Today, there are more than 100 million Z-Wave devices around the world~\cite{ZwaveStats}. The key reason for Z-Wave's success is its security, low cost and low power consumption, which make this technology suitable for sensors with small batteries. According to a recent study, Z-wave is used in 9 out of 10 leading security companies~\cite{ZwaveStats}. We examined 26 home security systems available on the market (such as Amazon's Ring, Hubitat, Lifeshield, Honeywell, ADT, etc.)  and found Z-Wave was used in 17 of them. 

\begin{figure}
    \centering
    \includegraphics[width=\columnwidth]{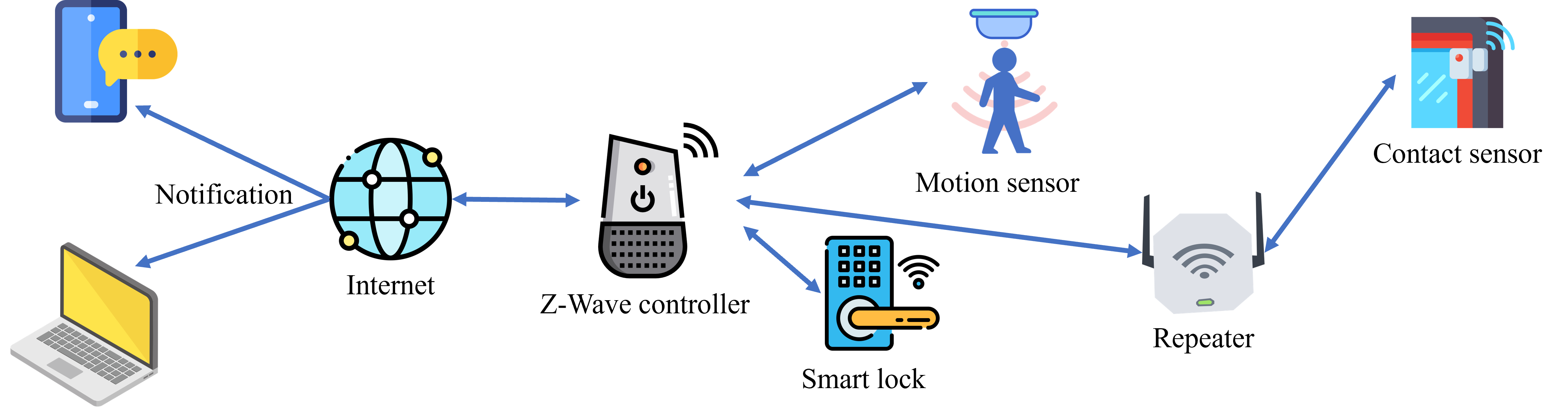}
    \caption{The topology of a Z-Wave network}
    \label{fig:topology}
\end{figure}

Z-Wave operates in the sub 1-GHz band and its exact frequency slightly varies in different countries. For example, Z-Wave devices communicate at 908-915 MHz in North America, and 868.42 MHz in Europe~\cite{Zwavefreq}. Z-Wave protocol consists of four main layers: the Physical Layer, the MAC layer, the Network Layer and the Application Layer~\cite{whatsYourProtocol}. The Physical Layer is responsible for modulation, choosing a channel and synchronization of a receiver through the use of a preamble. A Z-Wave preamble is a repeating sequence of 0s and 1s which act as a delimiter, allowing Z-Wave devices to identify unique messages. 
Figure~\ref{fig:topology} illustrates the topology of a Z-Wave network. A Z-Wave controller acts as a bridge to connect a variety of Z-Wave devices to the Internet. A Z-Wave network can include one or multiple repeaters to extend the coverage of this wireless network.

The MAC layer is responsible for collision avoidance, control of the medium between nodes in a Z-Wave network, and the MAC header. The MAC header holds important information such as the Home ID, Source ID, Destination ID, frame length and checksum. The Home ID is 32 bits and helps to identify which network a packet was sent on. The Source ID and Destination ID are each 8 bits and help to identify the sender and intended receiver of a packet on that network. The responsibilities of the Network Layer include maintaining a routing table, routing frames and scanning the topology of a network. Finally, the application layer provides services to the user using payloads sent or received. 

\begin{figure}[h]
\centering
    \subfigure[Wakeup Interval Devices]{
    \includegraphics[width=\columnwidth]{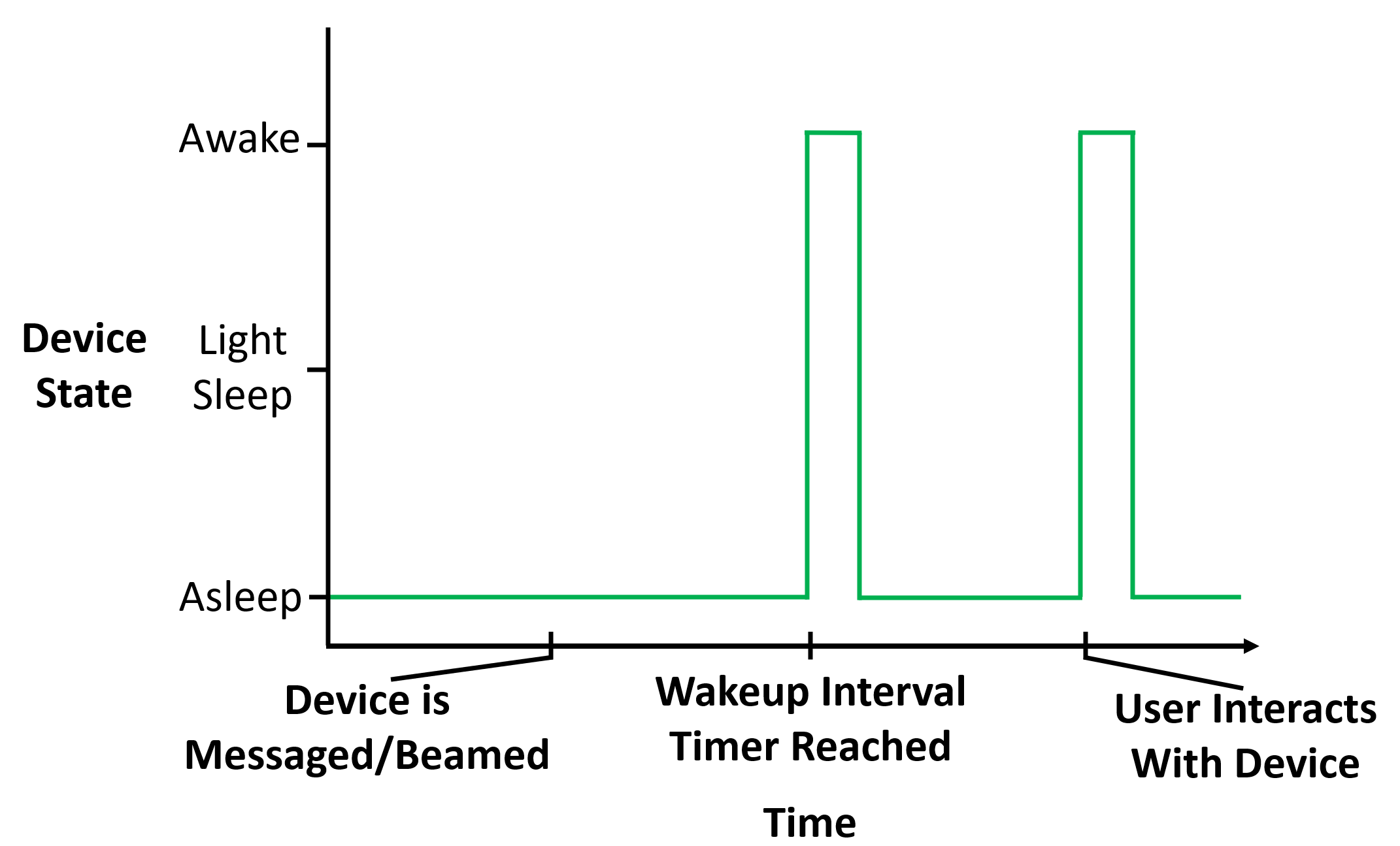}}
    \subfigure[FLiRS Devices]{
    \includegraphics[width=\columnwidth]{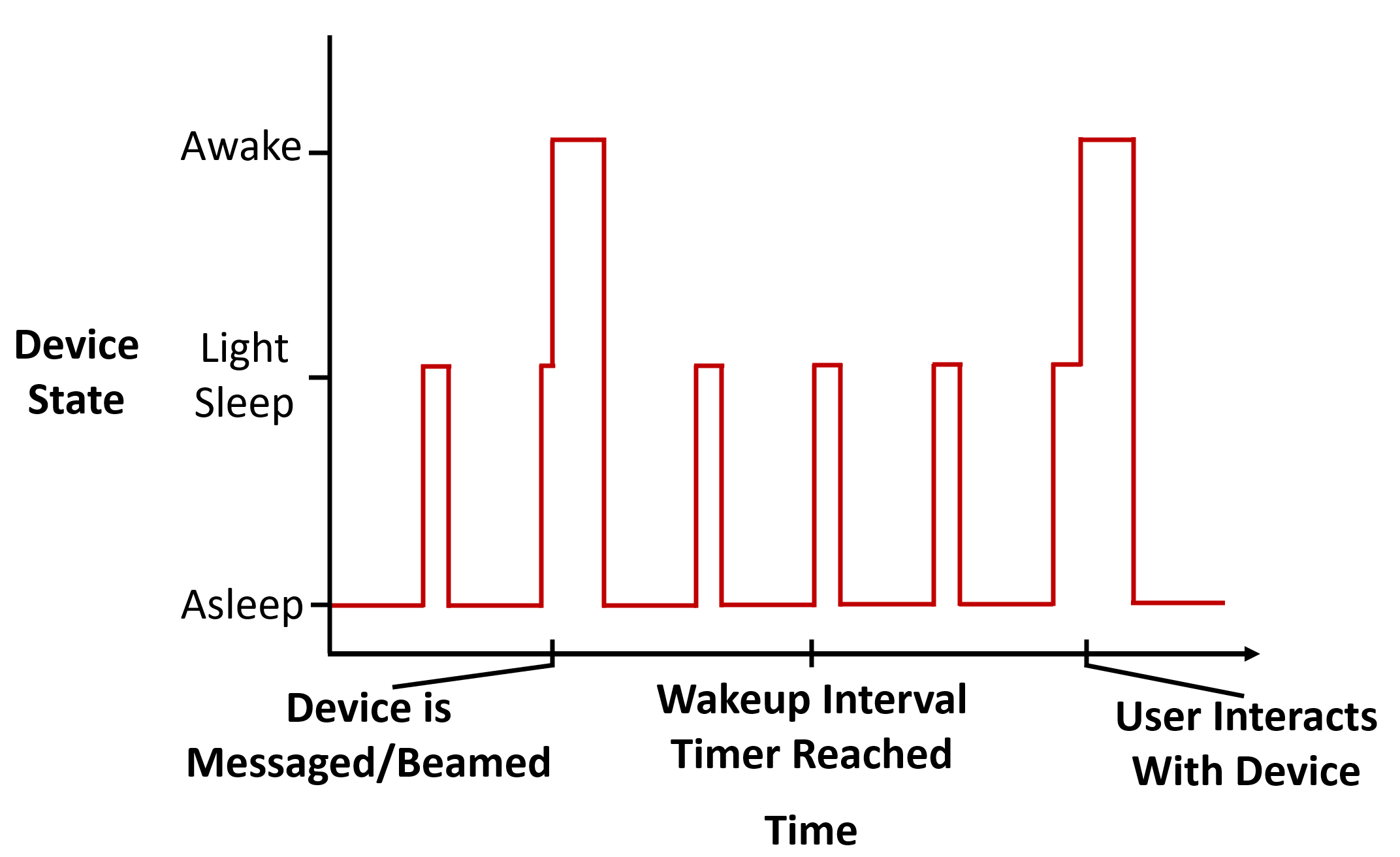}
    }
    \subfigure[Manual Wakeup Devices]{
    \includegraphics[width=\columnwidth]{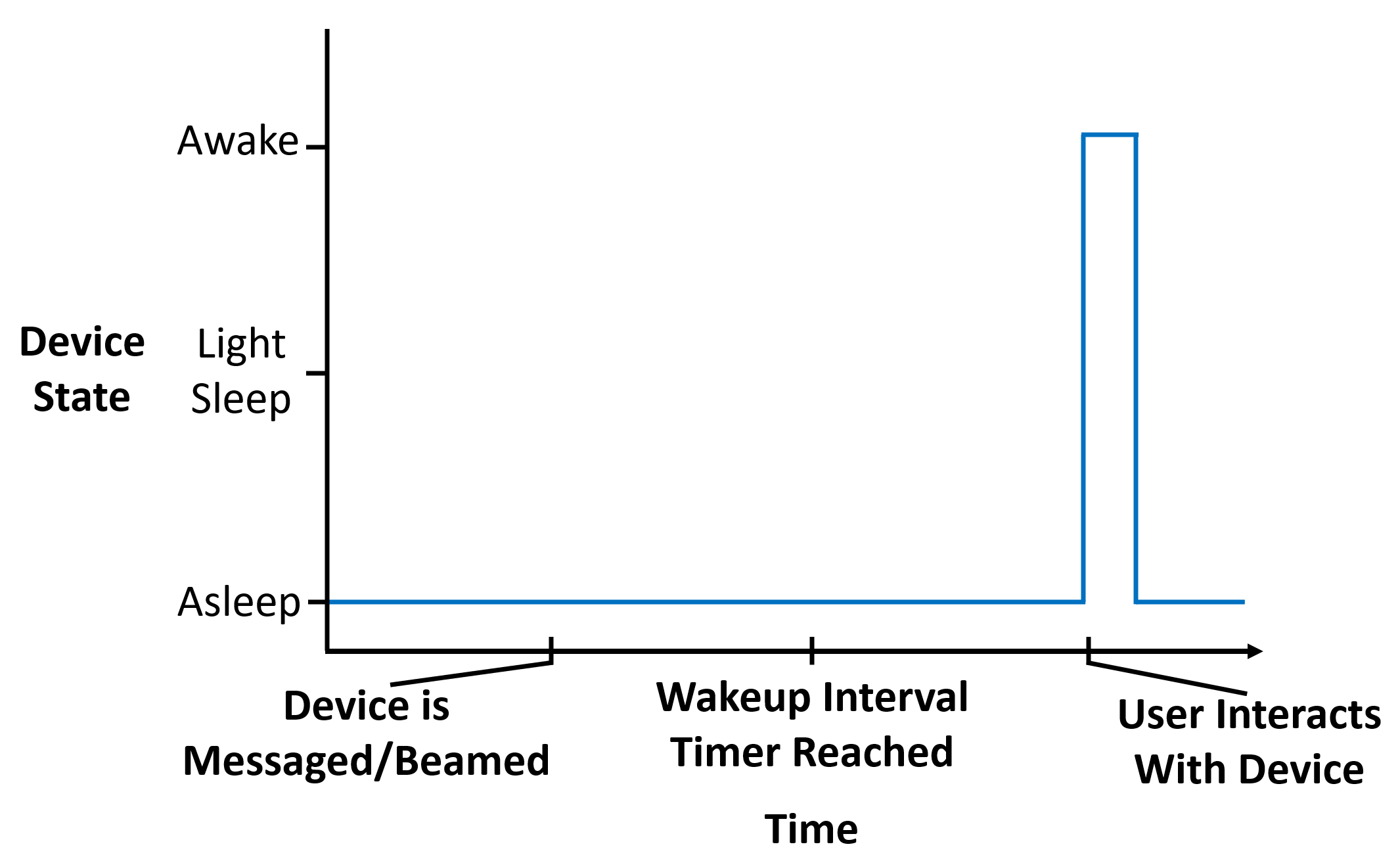}
    }
    \vspace{-15pt}
    \caption{Power State Changes in different Z-Wave devices}
    \vspace{-10pt}
    \label{fig:ZWaveTypes}
\end{figure}

\subsection{Z-Wave Power Saving}
Z-Wave devices are optimized to reduce battery consumption by alternating between an awake state and a sleep state. In the awake state, devices will listen for and transmit packets. However, while asleep, devices turn off their radio, and will not listen for or transmit any packet to conserve energy. 
This power saving mechanism allows Z-Wave devices to have a battery life of at least a few years in most devices.
As shown in Figure~\ref{fig:ZWaveTypes}, depending on the manner in which Z-Wave devices alternate between these two states, they are categorized into three groups: FLiRS devices, Wakeup Interval Devices and Manual Wakeup Devices~\cite{ZWaveUserManual}:

\textbf{Wakeup Interval Devices:} 
These devices have a timer that causes them to regularly wake up after a given amount of time. Once awake, these devices will announce that they have woken up by sending a wake up notification to the Z-Wave controller. The notification asks the controller to send any messages queued for the device. Once communication halts, the Wakeup Interval device will return to a deep sleep after a short amount of time. Common devices that exhibit this behavior are contact sensors and flood sensors. 

\textbf{FLiRS Devices:} 
Frequently Listening Receiver Slave (FLiRS)
 devices alternate between three power states. A fully awake state, a light sleep and a deep sleep. Similar to Wakeup Interval Devices, FLiRS devices will not send or receive any packet while in a deep sleep. However, at least once a second, they go to a lighter sleep for a brief amount of time. While in this lighter sleep, FLiRS devices will listen exclusively for a special message called a beam signal (transmitted by the controller). If a beam signal is detected, FLiRS devices will enter a fully awake state, where it will transmit and receive any packet. After a short amount of time in the fully awake state, the device will return back to a deep sleep. If the device is in a light sleep and no beam signal is received, the device will return back to a deep sleep. Some motion sensor devices belong to this category.

\textbf{Manual Wakeup Devices:} 
These Z-Wave devices alternate between a deep sleep and an awake state. 
These devices will remain in a deep sleep until they are used. Hence, they have the lowest power consumption, making them most suitable for applications such as car remotes, because the only time they should be powered is when they are used.

\begin{figure}[h]
    \centering
    \includegraphics[width=0.7\columnwidth]{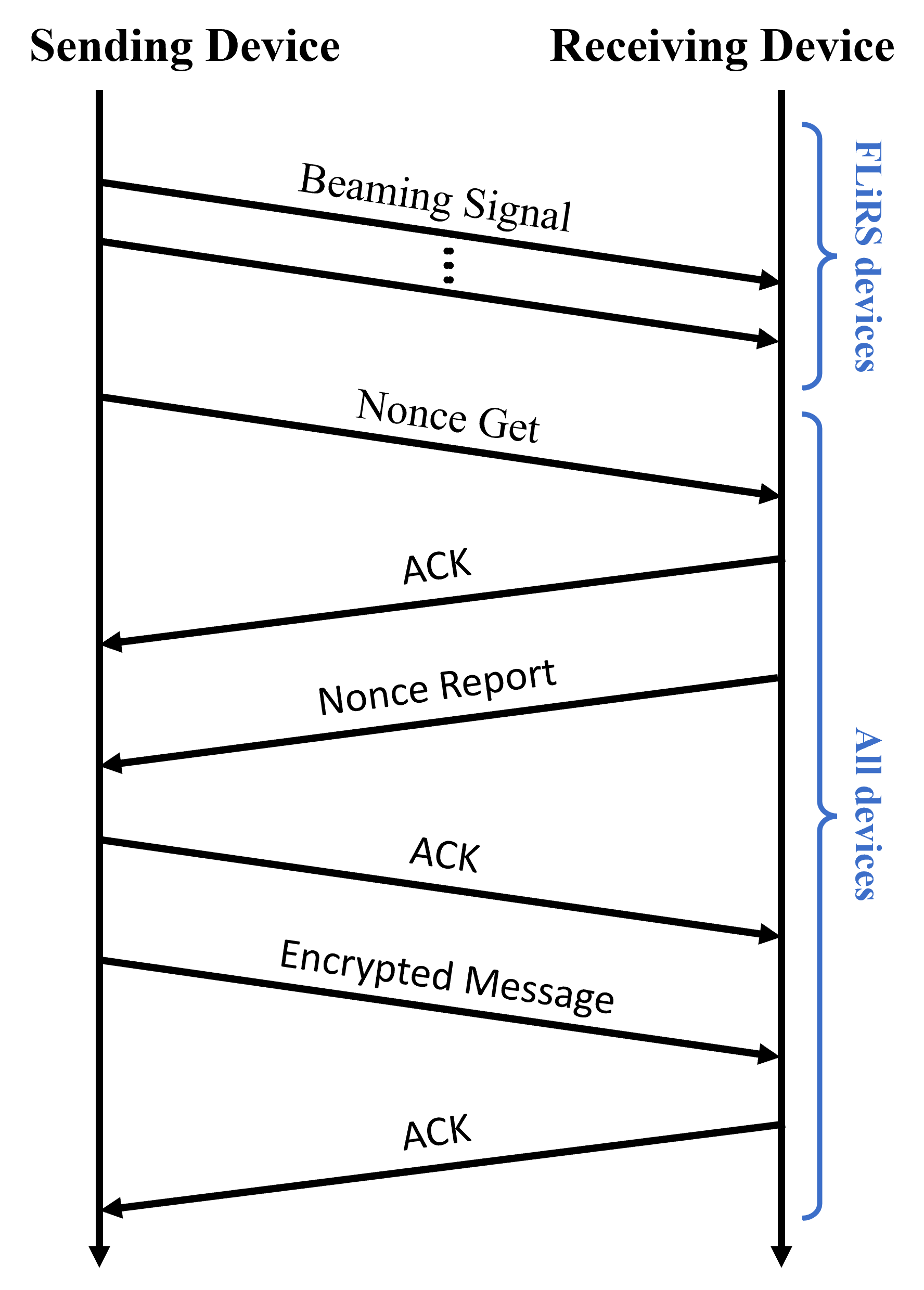}
    \vspace{-10pt}
    \caption{
    The Z-Wave protocol
    }
    \label{fig:protocol}
    \vspace{-15pt}
\end{figure}

\subsection{Z-Wave Frame Types}
We use different types of Z-Wave frames to execute attacks against Z-Wave devices. We now briefly review them here:

\textbf{Nonce-Get} messages are used to enable encrypted communication between a sender and a receiver as shown in Figure~\ref{fig:protocol}. First, a sender device sends a Nonce Get frame. When the receiver device receives a Nonce Get frame, it sends an Acknowledgment following with a Nonce Report. This Nonce Report frame contains a Nonce value, which is a unique and random number.
When the sender device receives the Nonce-Report command, it generates its own Nonce value. Using both the sender and receiver’s Nonce Values, the sender creates an Initialization Vector (IV).
The payload of the sender’s message can then be encrypted with the encryption key and the IV. 

A \textbf{Beam Signal} is a special frame sent to FLiRS devices to wake them as illustrated in Figure~\ref{fig:protocol}. 
Beam frames consist of only a preamble which is 8 or 20 bytes (including a one byte beam Tag, a one byte Node ID, and a one byte hashed Home ID). 
FLiRS devices wake up if they receive a beam signal during the light sleep period, otherwise they go back to sleep. Therefore, in order to talk to a FLiRS device we need to send some beam frames to it first.

\begin{figure}[t!]
    \centering
    \includegraphics[width=\columnwidth]{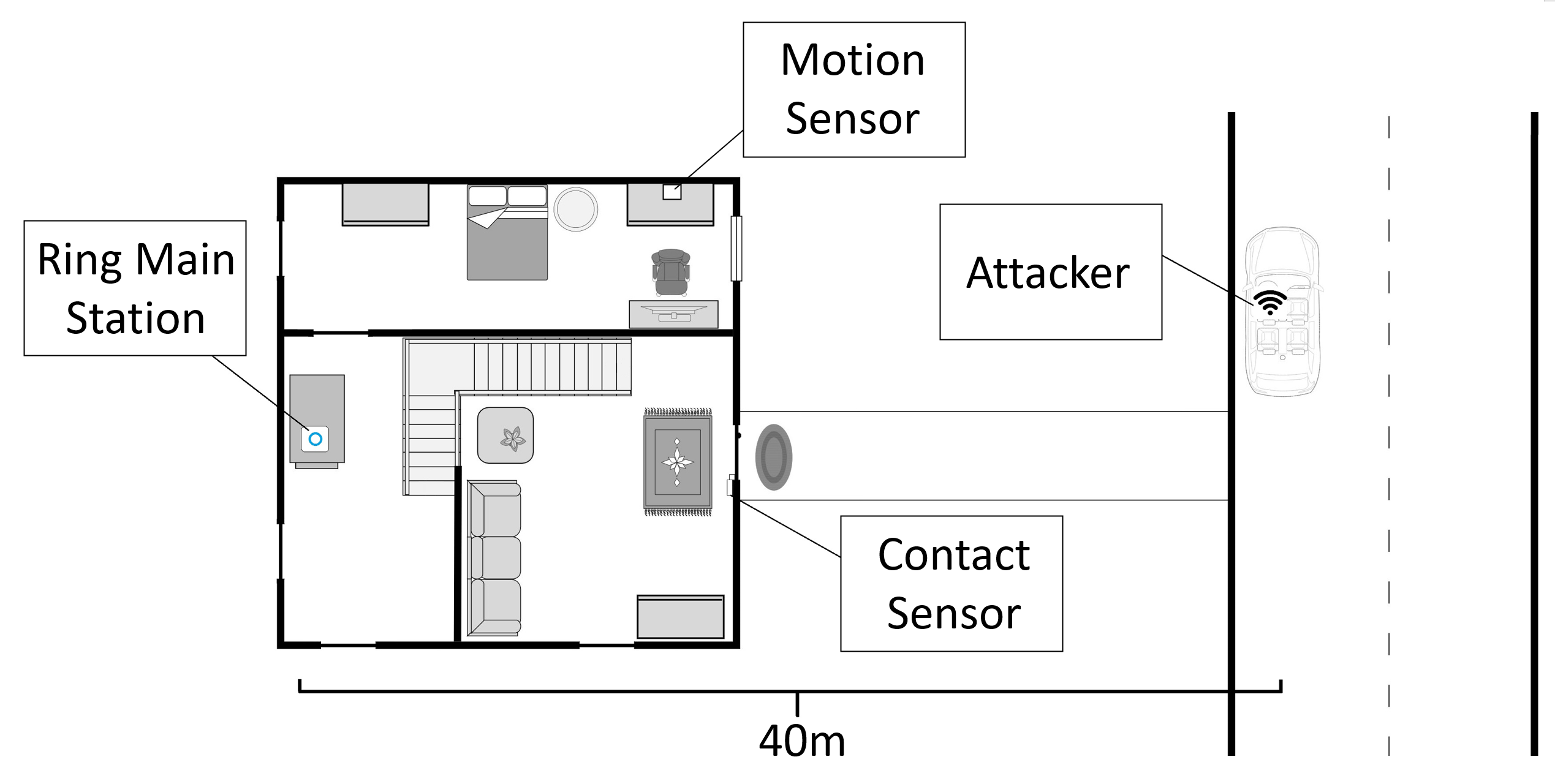}
    \caption{The attack setup. The Ring system is installed in a house and the attacker device is placed in a car 10s of meters away from the target house.}
    \label{fig:attacksetup}
\end{figure}

\section{Attack Setup}
We evaluate our battery drain and DoS attacks in a typical home equipped with a Ring Home Security System. We choose Ring's product in our setup since it is the most popular security system in the market. Moreover, the Z-Wave chipsets used in their sensors are the same as most other security systems and hence they have the same performance and flaws. In our setup, we target our attack to two common types of devices: The Ring contact sensor and Ring motion sensor. The contact sensor is a common sensor in security systems that monitors whether a door is open or closed. Contact sensors are a very important part of a security system as they are the front line defense of a home on the front and back doors and ground-level windows.

The contact sensor is a Wakeup Interval device with an interval of 12 hours, and a heartbeat delay of 71 minutes. The sensor uses two 235 mAh CR2032 batteries, giving the contact sensor a total capacity of 470 mAh. The manufacturer reports that each sensor has a three year battery life-span until new batteries are required. The Ring motion sensor is a FLiRS type device used to detect and report motion to the main station. The Ring motion sensor is a FLiRS device, and it uses two AA batteries. The manufacturer reports a three year battery life-span for this device. 

We use a Yard Stick One~\cite{yard-stick-one} as an attacker device. This device is available on Amazon and Adafruit, and costs around 100 USD. 
This off-the-shelf software defined radio allows us to generate and transmit arbitrary Z-Wave packets in the 900 MHz spectrum. The Yard Stick One was plugged into a 
laptop running Ubuntu 20.04.
We use this device to send packets to contact and motion sensors in our setup. During our experiments, all sensors are left completely untouched. 
We test our drain attacks for different distances between the attacker and target device. In particular, we test when the attacker is close to the target device and also when the attacker is in a car parked 40 meters from the targeted network, as shown in Figure~\ref{fig:attacksetup}. 
We also use a multimeter to measure the current consumption of the targeted device during our power measurement experiments. 
Note, although we use a laptop for our attacker device, our attack is so light in computation that one can simply implement it on a Raspberry Pi, powered by a USB battery bank. This allows the person to leave the attacker device close to the house without being around it, or put the attacking device in a car parked near the target building.

\section{Battery Drain Attack}
\label{sec:Z-waveBatteryAttack}
The main advantage of Z-Wave technology is its low power consumption which enables sensors to run on a small battery 
for many years. However, in this section, we show how an attacker can reduce the battery life of a Z-Wave device by orders of magnitude.
We find that an attacker can force a victim device to continuously transmit ACKs and Nonce Report messages by bombarding the victim device with Nonce Get messages.
This increases their power consumption significantly.
However, 
Z-Wave devices only Acknowledge Nonce Get packets from valid devices on their network. 
Interestingly, we find that an attacker device can do this by generating a packet that has the same Home ID, Source ID of another device in the network, such as the primary controller. Note, the Home ID, and Source ID of other devices can easily be found through sniffing, as they are not encrypted in Z-Wave protocol. In fact, we find that Z-Wave's access points typically have a Device ID of 0x01, making it a particularly easy device to impersonate for an attacker.

There is a technical challenge in executing this attack. Since Z-Wave devices are in sleep mode most of the time, how can the attacker start the attack? Moreover, even if a device is awake, this attack does not last long since Z-Wave devices go back to sleep to reduce their power consumption. 
Therefore, to completely kill a victim's battery, the attacker needs to use a technique to prevent the victim device from going back to sleep or wake it up as soon as it sleeps. 
Due to the differences in how FLiRS devices and Wakeup Interval devices transition from a deep sleep to an awake mode, different techniques must be utilized for each type.


\subsection{FLiRS Devices}
\begin{figure}[!bh]
    \centering
    \includegraphics[width=0.7\columnwidth]{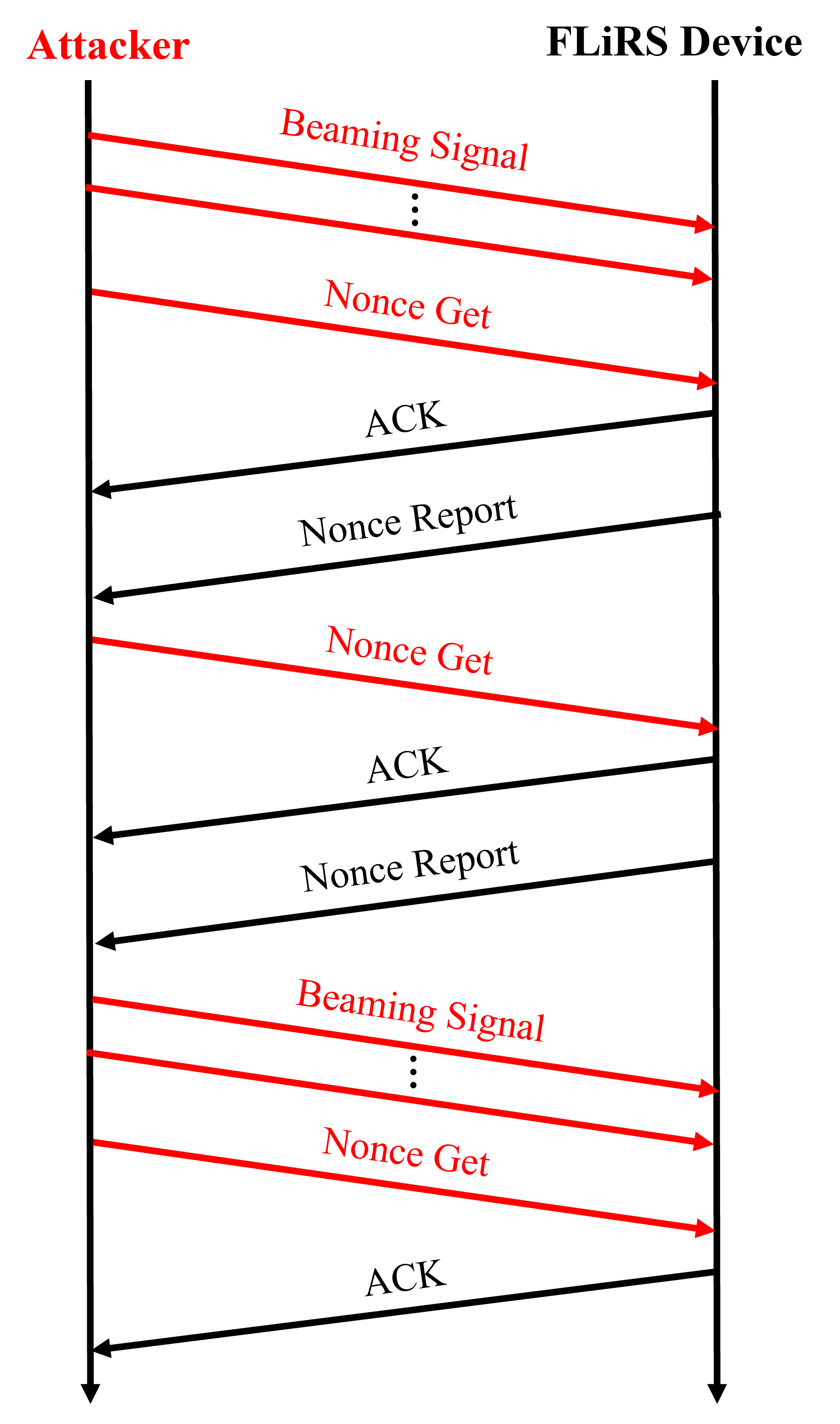}\\ 
    \caption{Battery drain attacks on FLiRS Z-Wave devices}
    \label{fig:protocolAttack-FLiRS}
\end{figure}

As explained in Section~\ref{sec:background}, FLiRS devices periodically go to a light sleep where they will listen for beam signal messages. If a beaming signal is detected, the device will enter the awake state. Therefore, we find that an attacker can periodically send the victim device a beam signal to wake it, followed by Nonce Get packets which impersonate another device on the network, as shown in Figure~\ref{fig:protocolAttack-FLiRS}. 

Finally, it is worth mentioning that although Z-Wave protocol recommends the beam signal last for 1160 ms to wake a device, we find that once a device is awake, a much shorter beaming signal (10 ms) would be enough to keep the device in the wake state, significantly increasing its power consumption.  

\subsection{Wakeup Interval Devices}
\begin{figure}[!bh]
    \centering
    \includegraphics[width=0.7\columnwidth]{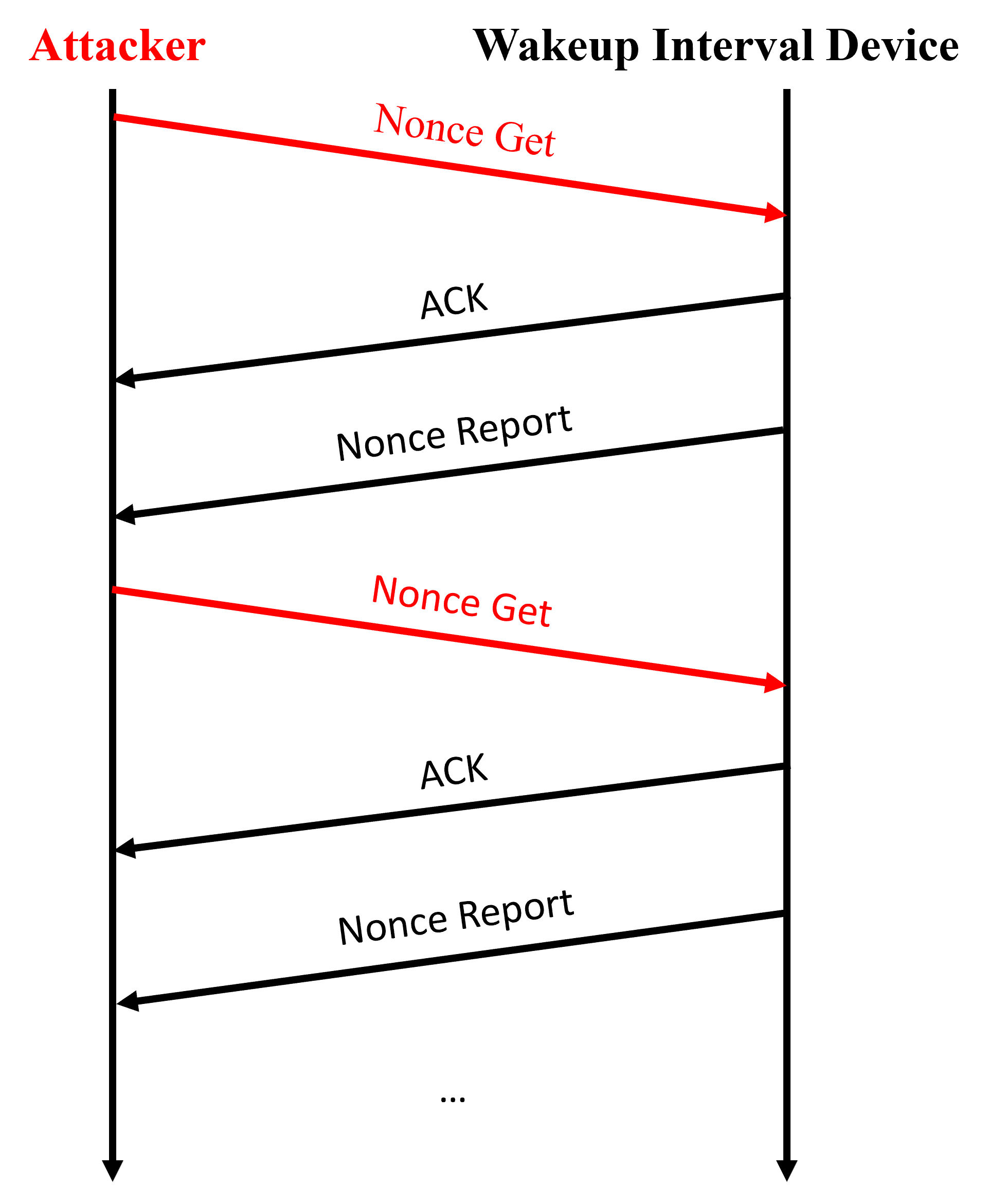}
    \caption{Battery drain attack on WakeupInterval  devices}
    \label{fig:protocolAttack-WI}
\end{figure}

Wakeup Interval devices cannot be forced to enter an awake state using beam signals. In particular, with Wakeup Interval devices, the battery drain attack can only begin once the device powers itself on, either due to an internal timer or external use. Note, the default Wakeup Interval on some sensors is very long. For example, the interval for the Ring contact sensor is 12 hours. However, we find that attackers can start their drains much faster. Interestingly, many Wakeup Interval devices such as the Ring contact sensor also have a mechanism called a Heartbeat. Heartbeats are battery reports which occur every 71 minutes. Heartbeats put the device in a powered state that can allow for a battery drain attack to begin. The packet exchange in this attack is illustrated in Figure\ref{fig:protocolAttack-WI}.
Moreover, any user activity (such as closing or opening a door) will wake the contact sensor. It is worth mentioning that through experimentation, we find that sending a minimum of three packets per second would be enough to keep the victim device in the awake mode and actively communicating.
Finally, we also find reducing the signal strength of the packets sent to the victim device appeared to further increase the power consumption of the victim device. This is due to the fact that the victim device needs to use a higher gain to amplify the received signal, resulting in an overall higher power consumption.

Figure~\ref{fig:z-wave-attack-code} shows a simplified version of the code we use to perform the battery drain attack on Wakeup Interval Devices attack. In addition to sending fake packets, another thread must check for responses transmitted by the target device to monitor the state of the target device. The code for attacking FLiRS devices is similar except that we need to send beam signals as well.

\begin{figure}[!h]
\begin{python}
from rflib import *

def createPacket(homeID, src, dst, payload):
    # Header (Preamble + Start of Frame Delimiter)
    d_init = "\x00\x0E"
    d_homeID = homeID.decode("hex")
    d_SrcNode = src.decode("hex")
    d_header = "\x41\x01"
    d_DstNode = dst.decode("hex")
    d_payload = payload.decode("hex")
    d_length = len(d_payload) + len(d_homeID) 
    + len(d_header) + 4
    d_length = format(d_length, '02x')
    d_length = d_length.decode("hex")
    d_checksum = checksum(d_init + d_homeID 
    + d_SrcNode + d_header + d_length + d_DstNode 
    + d_payload)
    return invert(d_homeID + d_SrcNode 
    + d_header + d_length + d_DstNode 
    + d_payload + d_checksum) 
    
#Defining the Nonce get packet
nonceGet = createPacket(homeID, src, dst, payload)

#Configuring the software radio
d = RfCat(0, debug=False)
d.setFreq(908420000)
d.setMdmModulation(MOD_2FSK)
d.setMdmSyncWord(0xaa0f)
d.setMdmDeviatn(20629.883)
d.setMdmChanSpc(199951.172)
d.setMdmChanBW(101562.5)
d.setMdmDRate(39970.4)
d.makePktFLEN(48)
d.setEnableMdmManchester(False)
d.setMdmSyncMode(SYNCM_CARRIER_15_of_16)
d.setMdmNumPreamble(MFMCFG1_NUM_PREAMBLE_8)
    
#Transmitting the fake Nonce get packet
while True:
    d.RFxmit(nonceGet)
    time.sleep(delay)

\end{python}
\caption{Python code for the battery draining attack on the contact sensor}

\label{fig:z-wave-attack-code}
\end{figure}

\begin{figure}[t!]
\centering
       \includegraphics[width=\columnwidth]{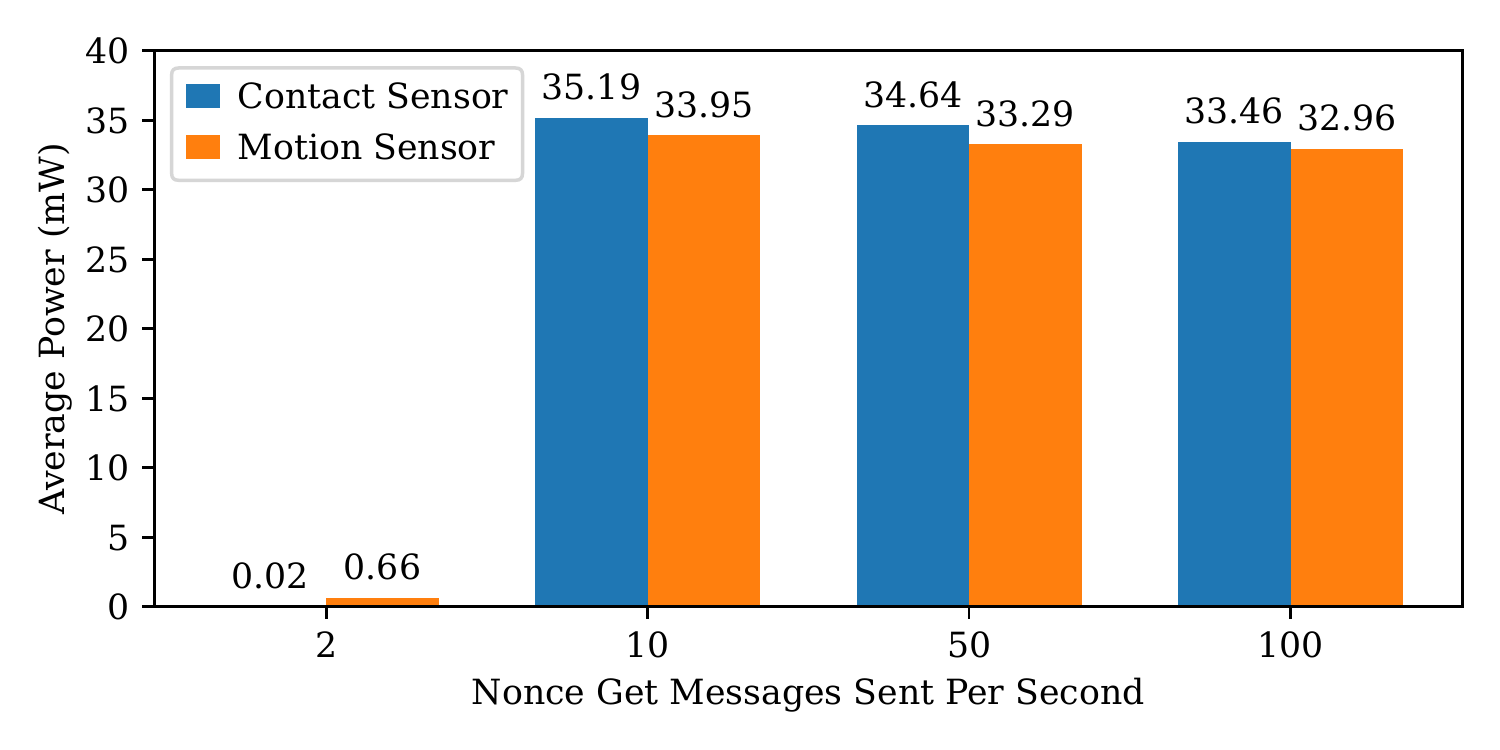}
    \vspace{-25pt}
    \caption{Power Consumption under draining attack
    }
    \label{fig:sensorpower}
    \vspace{-15pt}
\end{figure}

\subsection{Results}
\label{sec:Z-waveBatteryAttackResult}
We first measure the power consumption of the contact sensor in its normal deep sleep state. Our measurement shows that the sensor draws 0.02 mW during the sleep state. We then start our attack by sending 2 Nonce-Get packets per second (PPS) to the contact sensor. We notice the power consumption does not increase significantly as this quantity of Nonce Get messages is not enough to keep the device in a high power state. We then increase the number of packets to 10 PPS, 50 PPS and 100 PPS. Figure~\ref{fig:sensorpower} shows the current consumption for these scenarios. 
Our results show that when we send at least 10 packets per second to the victim, the power consumption substantially increases to 35 mW. This means that our attack increases the power consumption of the sensor by more than 1500 times.

We are able to perform this attack even when the attacker is 40 meters away from the target devices. Note, sending more than 10 packets per second does not increase the power consumption anymore. We believe that this is due to the fact that sending 10 packets per second was just enough to keep the sensor's radio awake most of the time. Moreover, in radios, receiver power consumption is in the same range as 
the transmitter's power consumption. 

Next, we continue our attack using 10 PPS until the battery is completely drained and the sensor does not work. We found that the life-span of a new battery is reduced to 16 hours from 3 years. In fact, this would be much shorter in reality since most sensor's batteries are not new. For example, if the battery has under 20\% of its energy left, the battery can be drained in a few hours. Note that this 20\% lasts for 7 months under normal usage.
In Section~\ref{sec:weakest-link}, we propose a heuristic that can be used by an attacker to find devices that have the lowest level of battery remaining.
An attacker can simply park a car 10s of meters away from the target house. An attacking device from inside the car drains the battery of a target device. Since no supervision is needed, 
the attacker can leave the car there for a while until the battery of the target device is completely drained. 

Next, we repeat the same attack and experiment on a motion sensor. We observe the same behavior. As shown in Figure ~\ref{fig:sensorpower}, the sensor consumes 0.65 mW during sleep mode or when the attacker is sending only 2 Nonce Get messages per second. However, our results show that when the attacker sends more than 10 packets per second, the power consumption is significantly increased, draining the battery.
Our attack increases the power consumption of the motion sensor by more than 51 times.

\subsection{Ramping Behavior} 
\begin{figure}[h]
\centering
    \subfigure[Number of messages during Ramping]{
    \includegraphics[width=0.9\columnwidth]{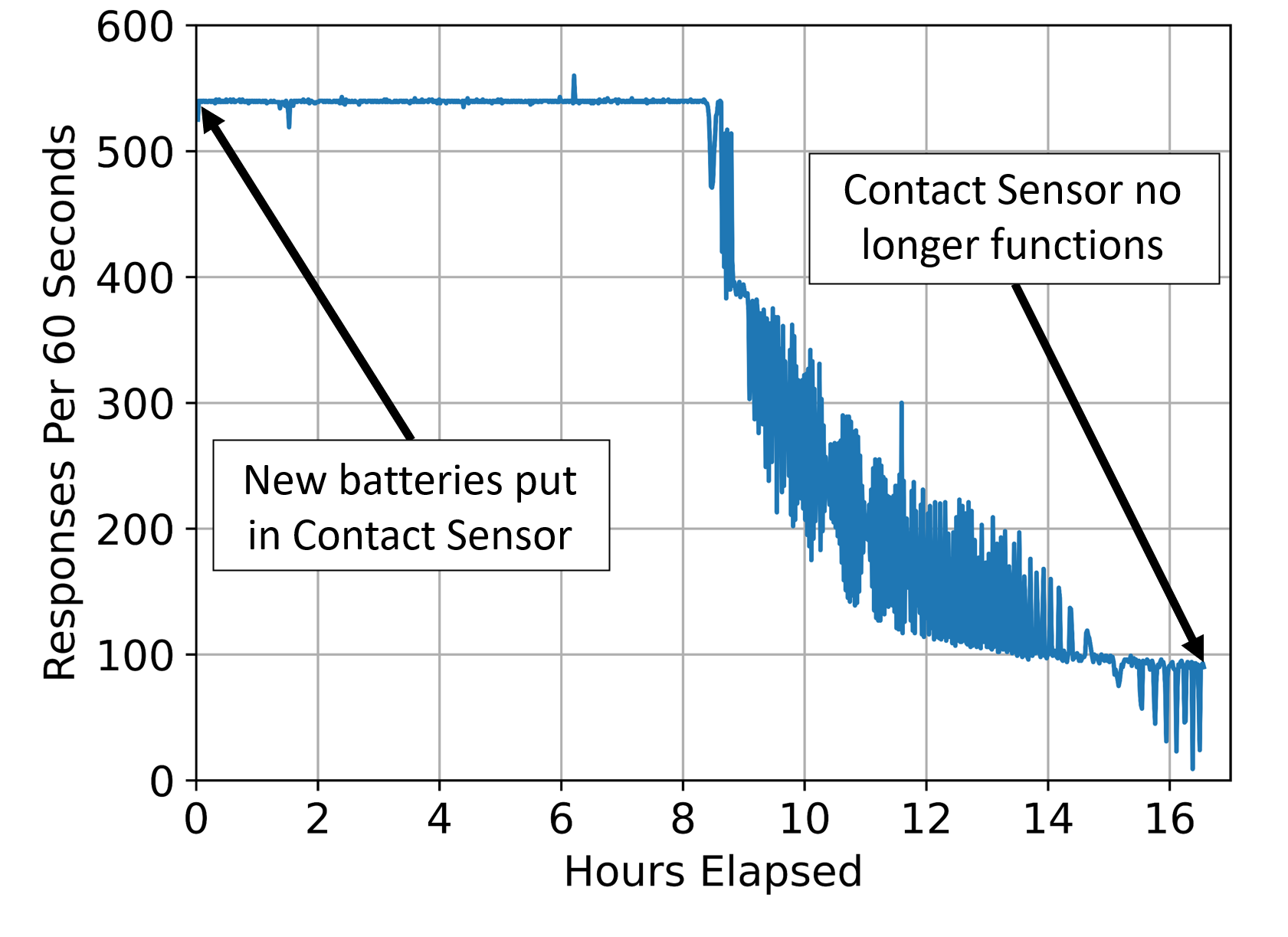}}
    \subfigure[Voltage during Ramping]{
    \includegraphics[width=0.9\columnwidth]{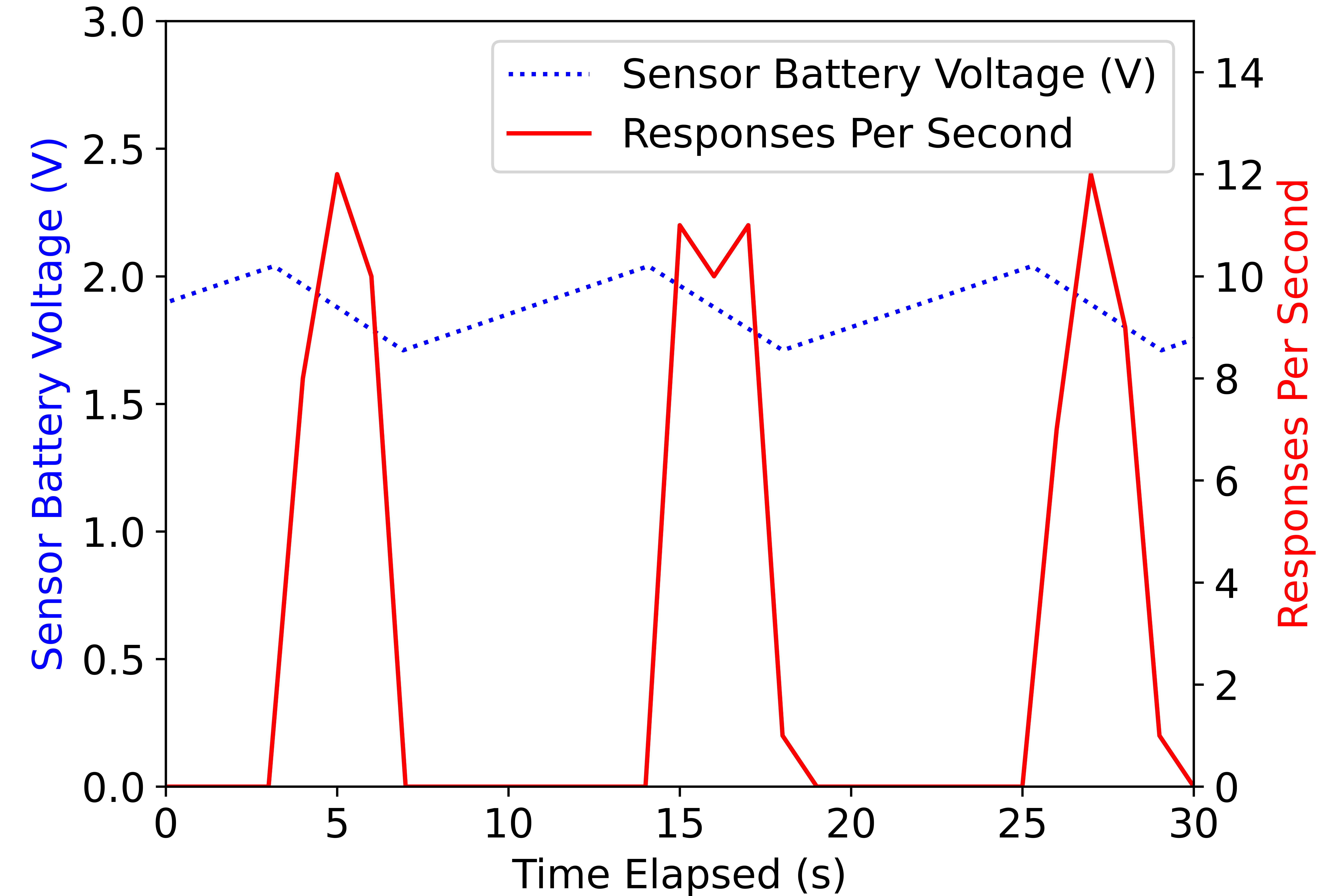}
    }
    \caption{The ramping behavior is when a Z-Wave device shuts down temporarily under attack.}
    \label{fig:ramping}
\end{figure}

During our battery drain experiment, we made an interesting observation.  We noticed as the battery drains, the average number of responses (received from the sensor) is reduced. Figure~\ref{fig:ramping}(a) shows this observation. In this experiment, we place a new battery in the contact sensor and we start the battery drain attack at time 0. Initially, the sensor sends around 550 responses per minute. However, approximately halfway through the drain, the  sensor begins to exhibit a behavior we call \emph{Ramping Behavior}.

The beginning of Ramping Behavior is characterized by a large drop in the average number of messages received from the target device. Although the attacker is still sending the same number of packets to the sensor, the average number of responses is reduced. One may assume that the contact sensor is simply sending messages at a slower rate, however, we found that the sensor is actually alternating between responding and  not responding. 


After investigation, it appears that the start of the break is due to a voltage drop in the battery caused by the battery drain. Figure~\ref{fig:ramping}(b) shows the battery voltage and the number of sensor responses per second, during 30 seconds of the attack while the contact sensor Ramps. As the sensor responds to the back-to-back packets sent by the attacker, the battery drops from 2V to 1.7V. At that point, the sensor can no longer operate since the voltage is too low, and hence it takes a break. However, after some time, the voltage gradually comes back to 2 V and the contact sensor turns back on, and the cycle repeats. As the life of the battery decreases, the faster the voltage drops and the longer the voltage takes to increase. 
Finally, at some point, the battery voltage never comes back to 2 V and that is when the battery is dead. 

\subsection{Which device to attack?}\label{sec:weakest-link}
\textit{``A chain is only as strong as its weakest link''}. 
This also applies to home security systems. For example, even if all windows and doors of a home are equipped with contact sensors, it only takes  one sensor to fail for the whole system to be compromised. 
Specifically, if one sensor is disabled, the attacker can enter the property from that door or window. Therefore, it does not matter if other sensors are still working.

Now, the question is how can an attacker find the weakest sensor (i.e., the one with the shortest battery life) in a home security system? To do so, our idea is that the attacker can use the length of the break period or quantity of responses received from the contact sensor during its Ramping Behavior to estimate the battery life of different sensors and attack the ones which have the lowest remaining battery capacity. In particular, the number of responses received can be a good indicator of a device's battery life. 


To test this idea, we conduct a series of  experiments in which we observe the behavior of a sensor with different battery levels under the battery drain attack. 
We first drain the battery of a sensor so that the Ramping behavior starts. Then we stop the attack for a few hours so that the battery can recover.
Then we start the attack again and measure the number of responses we receive from the sensor per minute. This experiment is indicated as Exp 1 in Figure~\ref{fig:ramping_different_levels}.
Because the battery has recovered to some extent the sensor steadily sends around 500 messages (i.e., no ramping behavior) when we start the experiment. 
However, after about 30 minutes the Ramping behavior starts. This is a strong indication for the attacker to infer that the battery of the target device has less than 50\% of its life left.

\begin{figure}[h]
    \centering
    \includegraphics[width=0.9\columnwidth]{figures/segmented.pdf}
    \caption{Ramping behavior with different levels of battery.}
    \label{fig:ramping_different_levels}
\end{figure}

After the first experiment reaches the Ramping behavior state, we once again stop the attack for a few hours and repeat the experiment three times until the battery is completely dead (i.e., EXP 2, 3, and4). Since we run these experiments back to back, the remaining battery at the start of each experiment is lower than the previous experiments.
We can see in the figure that as the battery loses its energy, the time needed to reach the Ramping behavior state decreases. 
There is a direct relationship between this time and the remaining battery.
An attacker can use this behavior to detect the sensor with the least battery life. In particular, the attacker can start the attack on all sensors in a house, and monitor the number of messages each device sends back. As soon as one device enters the Ramping behavior, the attacker can just continue the attack on that device because it will be the first device to die. Then the attacker can utilize the signal strength of the responses to figure out which window or door that specific sensor is installed on.

\section{Z-Wave Denial-of-Service Attack}
Besides the battery drain attack, we also find that Z-Wave devices are vulnerable to different Denial-of-Service (DoS) attacks that make a network, or parts thereof inaccessible. 
We show that these attacks are able to disable different types of Z-Wave devices instantaneously. For example, these attacks can disable a motion sensor so that if someone walks past the sensor nothing happens and no report is sent to the user of the security system.
In the following, we explain three different DoS attacks against Z-Wave devices.

\subsection{Denial of service to Z-Wave network} 
we find that if the attacker sends at least 50 packets per second to the Ring main station, the main station's functioning can be brought to a halt. Note, in these packets, the Destination Device ID is set to the Main Stations Device ID, which is normally equal to 01, the Source ID is set to the Device ID of some device on the network and the Home ID is set to the Home ID value of the network. An attacker can simply discover all these values by sniffing the Z-Wave signals. We have tried different types of frames and we found that the attack works regardless as long as the destination of the frame is the Z-Wave controller.

When the main station is denied service, the entire network is rendered useless making this attack particularly malicious. Finally, to confirm that the attack is a denial of service and the main station does not stop working because of occupancy of the frequency band, we run the same experiments when the Home ID is altered to be the ID of another network. We find that under these conditions, the Ring network resumed function. This finding confirms that this attack is a denial of service and is not a simple jamming attack. 
We found that the attack works even when the attacker is 100 meters away from the target Z-Wave network. This distance can be increased if the attacker uses a directional antenna.
    
\subsection{Denial of service to a motion sensor}
The next denial of service attack that we discovered is on the motion sensor. Denials of Service on the motion sensor are potentially dangerous as they could allow an attacker to easily move through a restricted area while evading detection. We find that if the attacker forces the motion sensor to remain awake through the use of beam signals, and then sends at least 80 messages (Configuration Get or Nonce Get messages) per second, the sensor stops sensing. 

We also find when the Home ID and Home ID Hash of the attacker's messages are changed to values not relating to the target network, the motion sensor is able to function, confirming that the failure of the motion sensor to operate is not due to occupying the frequency band. The denial of service of the motion sensor works up to 40 meters away from the device on the street.
    
\subsection{Denial of service to a contact sensor}
The final Denial of Service is on the contact sensor. As explained in ~\ref{sec:Z-waveBatteryAttackResult}, the battery drain attack takes around 16 hours to drain the Ring contact sensor. However, halfway through the drain, we notice a behavior which we call Ramping Behavior. As explained before, during this Ramping, the sensor alternates between responding and taking breaks which corresponds to the drop in the voltage of the device's battery.

We find that during these breaks, the contact sensor does not sense nor communicate. This means that although the full drain takes 16 hours, around the halfway point of the drain when the sensor is taking a break, an attacker can open a door with its associated contact sensor without setting off any house alarm. Given that the battery of user devices are usually around 20-60\% capacity, this DoS is particularly effective. We tried this attack at different distances. Our result shows that the denial of service of the contact sensor worked up to 40 meters away from the device on the street.

\section{Potential Defenses}
We now explore defenses against our proposed attack. Our insight is that the effectiveness of the attack mostly depends on the energy saving mechanism used in the sensors. Therefore, we propose different solutions which will make the energy saving mechanism smarter. 

The first solution involves changing when devices will return to a deep sleep. Our Z-Wave attack uses back-to-back Nonce Get messages to keep the target device in an awake state, resulting in draining their battery. To defend against this attack, we propose altering the power saving mechanism that puts the device back into a deep sleep if the device has been awake for a long time without receiving a meaningful packet. We define a meaningful packet as a packet which requires encryption. Therefore, since attackers cannot produce meaningful packets, they will have no way to keep a device in an awake state for an extended period of time, reducing the influence attackers have over that device. 

The second solution to defend against DoS attacks is to actively detect spoofing on the network. In this solution, Z-Wave devices need to listen to see if they receive any messages which contain their own Device ID as the Source ID value. Should a device receive a packet whose Source ID value is its own ID, it can immediately notify the base station that a malicious behavior is occurring. From this point, different actions can be taken depending on the system. For example, in a Ring Home Security System, the system should alert the user that malicious behavior is occurring. On other networks, the main base could send a special signal to a device to put it in a 'Suspicious State', where it will temporarily trust less to non encrypted messages. 

\section{Related Work}

A Z-Wave security flaw is presented in  \cite{ZWaveSecurityLocks} which allows attackers to take control of a victim's door locks. This attack is performed by exploiting a weakness in protocol allowing attackers to change the network key of a recently added device. Another related work has shown that an attacker can force a device to use less secure S0 security by abusing backwards compatibility~\cite{ZShave}. However, the main weakness of this line of work is that they all require the attacker to be present when the user is installing the devices since that is the only time the network key is publicly shared. In contrast, our attack does not have this requirement and it can attack devices at any time.

Different Denial of Service attacks have been proposed for Z-Wave devices~\cite{whatsYourProtocol, crushingZWave, zWaveBlackHole}. By changing the Source ID and Destination ID of a Nonce Get packet to the Device ID of a target device, these attacks would force the target device to send a Nonce Report to itself which in the presence of a routing Node results in a DOS attack. However, this DoS can be easily avoided if Z-Wave devices ignore packets with the same sender and destination ID. 
Finally, in \cite{ZWaveRogue,whatsYourProtocol} it was discovered that capturing and replaying HTTP requests allows attackers to send their own valid Z-Wave messages, letting them control devices on the network. However, these techniques require the attacker to know the admin ID and password of the associated router for a Z-Wave controller which is not practical.

\section{Conclusion}
This paper shows that the popular Z-Wave wireless technology used in most home security systems is not secure nor safe against attacks. Our results show that an attacker can remotely increase the power consumption of a security sensor by orders of magnitude, killing its battery in a matter of hours.
We also show how an attacker can execute DoS attacks, disabling these sensors for a period of time. 
We hope the research community pursues this line of research to both better understand the impact of these attacks and to develop effective defenses against them.

\bibliographystyle{ACM-Reference-Format}
\bibliography{main}

\end{document}